\documentclass[aps,prl,twocolumn,showpacs,groupedaddress,floatfix]{revtex4}  
\usepackage{graphicx}  
\usepackage{amssymb}   
\usepackage{filecontents}
\usepackage{amsmath}
\begin{document}

\title{Non-adiabatic ponderomotive effects in photoemission from nanotips in intense mid-infrared laser fields}
\author{J. Sch\"{o}tz$^{1,2}$}
\author{S. Mitra$^{1,2}$}
\author{H. Fuest$^{1,2}$}
\author{M. Neuhaus$^{2}$}
\author{W.A. Okell$^1$}
\author{M. F\"{o}rster$^3$}
\author{T. Paschen$^3$}
\author{M.F. Ciappina$^{1,4}$}
\author{H. Yanagisawa$^{1,2}$}
\author{P. Wnuk$^{1,2}$}
\author{P. Hommelhoff$^3$}
\author{M.F. Kling$^{1,2}$}

\affiliation{$^1$Max Planck Institute of Quantum Optics, Hans-Kopfermann-Str. 1, D-85748 Garching, Germany}
\affiliation{$^2$Department of Physics, Ludwig-Maximilians-Universit\"{a}t Munich, Am Coulombwall 1, D-85748 Garching, Germany}
\affiliation{$^3$Department of Physics, Friedrich-Alexander-Universit\"{a}t Erlangen-N\"{u}rnberg, Staudtstra{\ss}e 1, D-91058 Erlangen, Germany}
\affiliation{$^4$Institute of Physics of the ASCR, ELI-Beamlines, Na Slovance 2, 182 21 Prague, Czech Republic}

\date{\today}

\begin{abstract}
Transient near-fields around metallic nanotips drive many applications, including the generation of ultrafast electron pulses and their use in electron microscopy. We have investigated the electron emission from a gold nanotip driven by mid-infrared few-cycle laser pulses. We identify a low-energy peak in the kinetic energy spectrum and study its shift to higher energies with increasing laser intensities from $1.7$ to $3.7\cdot10^{11} \mathrm{W}/\mathrm{cm}^2$. The experimental observation of the upshift of the low-energy peak is compared to a simple model and numerical simulations, which show that the decay of the near-field on a nanometer scale results in non-adiabatic transfer of the ponderomotive potential to the kinetic energy of emitted electrons and in turn to a shift of the peak. We derive an analytic expression for the non-adiabatic ponderomotive shift, which, after the previously found quenching of the quiver motion, completes the understanding of the role of inhomogeneous fields in strong-field photoemission from nanostructures.
\end{abstract}

\maketitle

\section{Introduction}
The evanescent electromagnetic near-fields around nanostructures lead to field enhancement and confinement down to a few nanometers, well below the wavelength of the exciting optical light \cite{Novotny06}. Exploring and exploiting these effects has led to the development of nano-optics, including plasmonic nanofocusing \cite{Stockman04,Ropers07}, ultrafast multi-dimensional nanoscopy \cite{Aeschlimann11}, the generation of extreme ultraviolet radiation with plasmonic waveguides \cite{Han16}, and time-resolved spectroscopy on the nanoscale \cite{Raschke12}. Femtosecond nanometer-sized electron sources have been realized with metal nanotips via nonlinear photoemission \cite{Hommelhoff06_1,Batelaan07,Ropers07,Schenk10,Ropers10}, and found applications in electron microscopy \cite{Barwick13,Ropers14,Ernstorfer14,Lienau15}. The sub-femtosecond control of the electron emission in strong fields was demonstrated for nanoparticles, nanotips and nanowire tips \cite{Zherebtsov11,Krueger11,Krueger2012,Wachter2012,Herink12, Lienau14,Suessmann15, Foerg16,Rupp17, Ahn17,Hommelhoff15, Ciappina17}. \newline
In strong-field photoemission an important parameter is the ponderomotive potential $U_p = e^2 \mathcal{E}_0^2 / (4m \omega^2)$, the kinetic energy of an electron due to its’ quiver motion in an oscillating field of amplitude $\mathcal{E}_0$ and frequency $\omega$ ($e$ and $m$ are the electron's charge and mass, respectively). Strong-field photoemission has first been observed from gas atoms \cite{Agostini1979}, where the electron dynamics can typically be separated into sub-cycle and drift motion \cite{Walther2002,Eberly1988,Corkum93}. The sub-cycle motion occurs on the timescale of the laser cycle after electron emission, where electrons undergo a laser-driven quiver motion, and can rescatter with the parent ion. Since the extend of electron motion within one optical cycle is typically much smaller than the focal spot size of the driving laser light, the electrons experience a quasi-homogeneous field on this time- and length-scale. On a cycle-integrated scale, i.e. for the remainder of the laser pulse after photoemission, electrons may experience some field inhomogeneity as they are drifting out of the laser focus. Here, the ponderomotive potential is adiabatically transferred to kinetic energy of the electron’s drift motion \cite{Kruit83}. For femtosecond laser pulses, the electromagnetic field typically switches off before electrons have moved out of the focus \cite{Bucksbaum87} and ponderomotive shifts can therefore be neglected.\newline
For nanostructures, at sufficiently high intensities, photoemitted electrons can experience the near-field decay of the enhanced field already within the sub-cycle evolution of the laser field. It has been shown that this strongly modifies the sub-cycle dynamics leading to a quenching of the quiver motion and suppression of electron rescattering \cite{Herink12}. An adiabaticity parameter $\delta=l_f/l_q$ has been introduced \cite{Herink12}, where $l_f$ is the near-field decay length and $l_q=e \omega^2 \mathcal{E}_0 /m$ is the amplitude of the quiver motion. However, the influence of the near-field decay on the electron drift motion has not been discussed.\newline
We investigate strong-field photoemission from a gold nanotip and observe a low-energy peak and its energy up-shift with intensity. With the help of a simple model and semi-classical Monte-Carlo trajectory simulations, we show that sub-cycle electron dynamics and drift motion become interdependent. The relevant mechanism for the up-shift is identified as non-adiabatic ponderomotive shift, which is important for small near-field decay-length and all pulse lengths, even few-cycle laser excitation. Based on the results, we introduce an analytic expression for the non-adiabatic ponderomotive shift. The results complete our understanding of the effect of inhomogeneous near-fields in photoemission from nanostructures.

\section{Experimental Setup}

\begin{figure}[htbp]
	\centering
	\includegraphics[width=\linewidth]{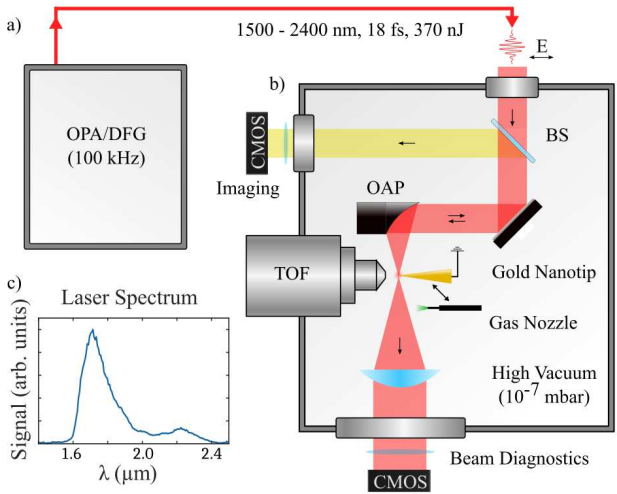}
	\caption{Experimental setup: a) NOPA-DFG-setup, which is seeded by a multipass-CPA-Innoslab-amplifier. b) Vacuum chamber hosting the experimental setup for the
		photoemission experiments along with an imaging setup used for coarse optimization of tip position and focus size. c) Spectrum of the laser pulses used in the experiments.}
	\label{fig:exp_setup}
\end{figure}

The laser system is based on non-collinear optical parametric amplification (NOPA) and subsequent difference frequency generation (DFG) (Fig.\,\ref{fig:exp_setup}\,a), details will be published elsewhere\cite{Neuhaus_tobepublished}). Briefly, the laser is based on a Yb:YAG laser delivering 1.6\,ps pulses centered at 1030\,nm operating at 100\,kHz. A first part is split off and used for white light generation in a YAG crystal and a second part is frequency doubled. The former serves as a seed for the NOPA stage pumped by the latter.  A third part seeds the DFG stage, which is pumped by the output of the NOPA. After compression pulses centered at 1800\,nm (see spectrum in Fig.\,\ref{fig:exp_setup}\,c)) with 370\,nJ and a duration of 18\,fs, as measured by a frequency-resolved-optical-gating (FROG) setup, are delivered to the experiment. The pulses are passively phase stable but not actively stabilized, which leads to a slow drift of 2$\pi$ over the course of around ten minutes, as checked by an f-to-2f-interferometer. The acquisition time of a single spectrum is shorter than the CEP-drift, such that the spectra are effectively for a small CEP range. The pulse energy is controlled by a neutral density filter wheel, which reflects a variable fraction of the power.

The experimental setup is shown in Fig.\,\ref{fig:exp_setup}\,b) and consists of a vacuum chamber which is pumped by turbo-molecular pumps to a base pressure of around $10^{-7}$\,mbar. The laser is focused by an off-axis parabolic mirror (OAP, f=15\,mm) to a spotsize of 2.6\,$\mu$m as measured by a knife edge setup. The nanotip and gas nozzle are mounted on an xyz-translation stage. The nanotip is electrically grounded. The laser is linearly polarized along the tip orientation which is parallel to the axis of the time-of-flight spectrometer (TOF, Kaesdorf ETF10). The electrons are detected on a microchannel-plate at the back of the TOF. In our experimental configuration the spectrometer has an acceptance angle of $7^{\circ}$.  The electrons TOF is registered by a time-to-digital converter (FAST ComTec P7889) and subsequently converted to kinetic energy.
The nanotip is positioned using an image of the focus by a refocusing lens. Additionally, it proved helpful to use the OAP itself for imaging by introducing a beamsplitter (BS) which reflects light (external illumination, reflected laser or stray light) originating from the focus and collimated by the OAP to an imaging stage outside the vacuum chamber. Fine alignment is done by maximizing the electron countrate in the spectrometer.
By removing the nanotip and driving the gas nozzle close to the focus, ATI spectra from Xe atoms are recorded, which serve as an intensity calibration by observing the well defined direct electron emission cutoff. Furthermore, by observing equally spaced photon peaks down to 1 eV, we can conclude that our system is well calibrated even for low-energy electrons.


\section{Experimental Results}

\begin{figure}
	\centering
	\includegraphics[width=\linewidth]{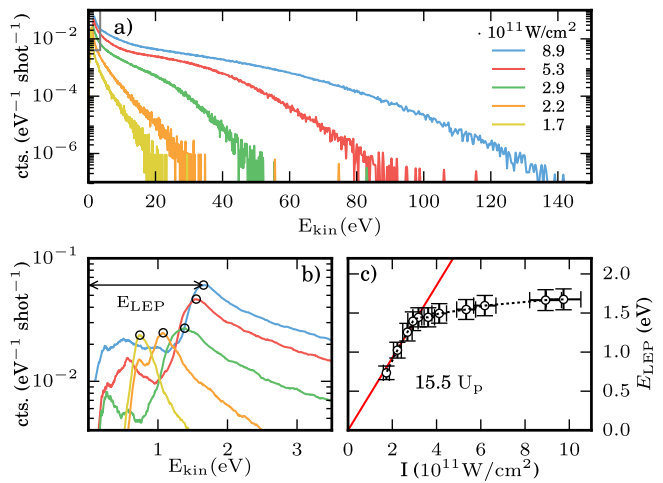}
	\caption{  a) Experimental spectra from a gold nanotip under irradiation for the indicated incident intensities. b) Low-energy region of the above spectra, dots indicate LEP positions $\mathrm{E}_{\mathrm{LEP}}$. c) LEP position versus incident intensity. For low intensities an approximately linear dependence with a slope of $15.5\,\mathrm{U}_{\mathrm{p}}$ is found (solid line).}
	\label{fig:exp_spectra}
\end{figure}

Figure\,\ref{fig:exp_spectra}\,a) shows the measured kinetic energy spectra for electrons emitted from a sharp gold nanotip for various incident intensities. The spectra are dominated by a low-energy peak (LEP). For the lowest intensities a near-exponential decay of the spectra is observed, while for higher intensities a high-energy shoulder is formed. The highest observed electron energies scale roughly linear with intensity and are significantly higher than expected from the input intensities, indicating the occurrence of field enhancement at the nanotip. Figure\,\ref{fig:exp_spectra}\,b) shows the enlarged low-energy region, resolving the LEP positions as well as smaller secondary peaks at lower energies (discussed later). With increasing intensity, the LEP value $\mathrm{E}_{\mathrm{LEP}}$ shifts to higher values.\newline
Fig.\,\ref{fig:exp_spectra}\,c) shows the intensity-dependence of the LEP position. With increasing intensity, the peaks shift to higher energies. We find that for small intensities the LEP position is approximately linearly increasing with a slope of roughly 15.5\,$\mathrm{U}_\mathrm{p}$ (solid line). Beyond intensities of around $3\cdot10^{11} \mathrm{W}/\mathrm{cm}^2$, the slope continuously decreases and the LEP positions are not directly proportional to the ponderomotive potential anymore. In order to understand this nonlinear behaviour, we will reexamine the electron dynamics in inhomogeneous fields in the next section. Note, the range over which data points can be obtained is limited by countrate on one hand and laser-induced damage of the nanotip on the other hand. \newline
Experimentally, less than 0.4 electrons per shot are detected at the highest intensities. Considering the detection efficiency of our setup, we can conclude that even at the highest
intensity not significantly more than one electron per shot is emitted, such that multi-electron interaction should not play a major role. Moreover, previous work carried out on strong-field photoemission, showed that space-charge effects lead to a shift towards lower energies \cite{Velasquez_Garcia14,Yanagisawa16} and even suppression \cite{Zherebtsov11,Suessmann15} of the LEP, which is in contrast to our observations.

\section{Theory}
\begin{figure}[!htbp]
	\centering
	\includegraphics[width=\linewidth]{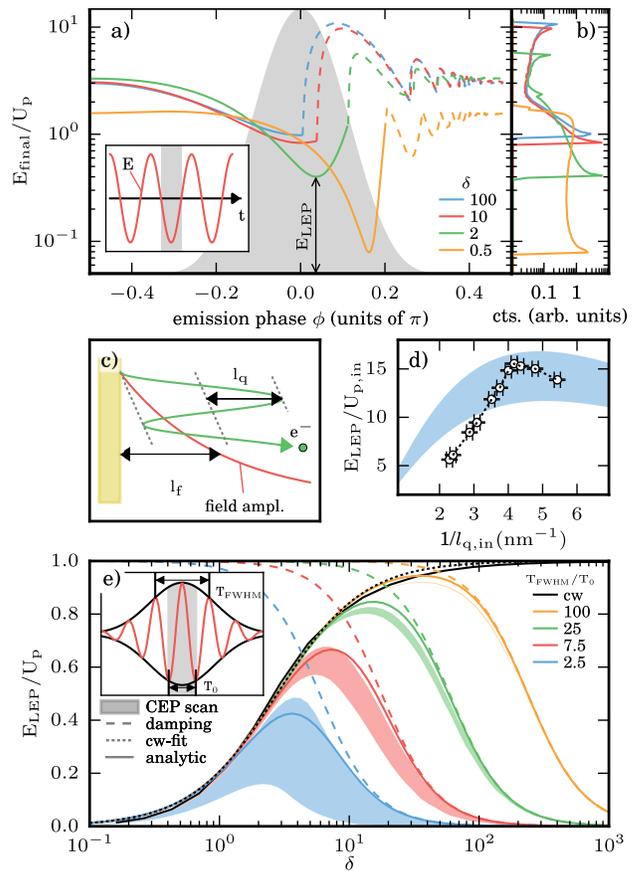}
	\caption{a) The final kinetic energy for different values of the adiabaticity parameter for a cw laser in terms of the oscillation period $T_0$ and $U_p$. Direct (rescattered) electrons are indicated as solid (dashed) lines. The gray area shows the emission probability according to Fowler-Nordheim tunneling (work function 5.5\,eV, intensity $\mathrm{3}\cdot \mathrm{10}^\mathrm{13}\mathrm{W/cm}^2$) on a linear scale from 0 to 1. 
		b) Electron spectra resulting from weighting the final energies with the emission probability. Clear low-energy peaks appear, whose position is given by the minima in the left plot.
		c) The trajectory of a direct electron illustrating the definition of the quiver amplitude $\mathrm{l}_\mathrm{q}$ and field decay length $\mathrm{l}_\mathrm{f}$.
		 d) Low-energy region of the above spectra, dots indicate LEP positions $\mathrm{E}_{\mathrm{LEP}}$. c) $\mathrm{E}_{\mathrm{LEP}}$ normalized by the ponderomotive potential against the inverse quiver amplitude ($\propto \delta$) together with the expectations from the analytic expression (shaded area) for a field decay length of 4.5$\pm$0.5nm and a field enhancement factor of 5.5$\pm$0.5.
		c) Dependence of the lowest final kinetic energy of electrons emitted during the central cycle of a Gaussian pulse on the adiabaticity parameter for
		different pulse lengths for all CEPs (shaded areas) together with $E_{\mathrm{shift,p}}/U_{\mathrm{p}}$ (dashed lines), the heuristic fit to the cw curve (dotted line) and the analytic expression (solid lines) as described in the text.}
	\label{fig:1D-sim}
\end{figure}
In order to investigate strong-field photoemission in inhomogeneous fields, we use the Simple-Man's-Model (SMM)\cite{Corkum93} adapted to inhomogeneous fields. Here, the electron is treated as a classical particle traveling in one dimension, influenced by a time-varying
electric field $\mathcal{E}(x,t)$ representing the near-field at position $x$ and time $t$.
For simplicity, we assume an exponential decay of the electric near-field with decay length $l_f$: $\mathcal{E}(x,t)=\mathcal{E}_0 \cdot \exp(-x/l_f)\cdot f(t)$, where $\mathcal{E}_0$ is the electric field amplitude
and $f(t)$ is the normalized temporal shape of the near-field. Introducing the dimensionless variables $\tilde{x}=x/l_f$ and $\tilde{t}=t/T_0$, where $T_0$ is the
oscillation period of the electric field, we arrive at the dimensionless equation (see \cite{Schoetz17}):
\begin{equation}
\label{dimless_equation}
\frac{\mathrm{d}\tilde{v}}{\mathrm{d}\tilde{t}}=-\frac{e T_0^2}{m l_f}\mathcal{E}_0\cdot f(\tilde{t}) \cdot \mathrm{e}^{-\tilde{x}}=-\frac{4 \pi}{\delta} f(\tilde{t})\cdot \mathrm{e}^{-\tilde{x}}
\end{equation}
This equation shows that in the case of inhomogeneous fields, the dynamics is governed by the adiabaticity parameter $\delta$, which indicates how much of the field inhomogenity the electrons experience during one oscillation of the electromagnetic field. In order to describe the electron motion, Eq. (1) is integrated numerically. The electron propagation is started at the surface with zero initial velocity. A reflection coefficient of unity for electrons recolliding with the surface of the nanostructure has been assumed.\newline
We first study the effect of the field inhomogeneity on the kinetic energy of the emitted electrons in continuous wave (cw) laser fields [$f(\tilde{t})=-\mathrm{cos}(2\pi\cdot \tilde{t})$]. Figure\,\ref{fig:1D-sim}\,a) shows final electron energies depending on the emission phase $\phi=2\pi\cdot \tilde{t}$  normalized to the ponderomotive potential $U_\mathrm{p}$ at the surface, for varying $\delta$.
For large $\delta$, electrons do not experience the inhomogeneity of the laser field on few-cycle timescales. Electrons emitted before $\phi=0$ do not rescatter and correspond to direct electrons (solid lines). Electrons emitted between $\phi=0~\mathrm{to}~0.26\cdot \pi$ rescatter once and form a distinct peak and can gain higher energies. Consecutive peaks form with electrons undergoing an increasing number of collisions (dashed lines). As the electrons drift out of the near-field, on a cycle-integrated timescale, they are accelerated by the gradient of the ponderomotive potential and their final energies are thus up-shifted by the ponderomotive potential $U_p$\cite{Eberly1988}.\newline
For decreasing $\delta$, electrons start to experience the field inhomogeneity already during their sub-cycle propagation. They are driven so far from the surface within one half-cycle that the field strength of the second half-cycle, which drives them back to the surface, is decreased. The reduction of $\delta$ leads to a decrease of the final energies of direct and rescattered electrons as can be seen from the maximum energy of direct and rescattered electrons in Fig.\,\ref{fig:1D-sim}\,a). This is a consequence of changed electron dynamics on sub-cycle timescales and has been studied experimentally before\cite{Herink12}. Yet, another very important feature, the reduction of the minimum energy (which determines the LEP as shown in Fig.\,\ref{fig:1D-sim}\,b) has so far been overseen. For decreasing $\delta$ drift and quiver motion can not be separated anymore. This is illustrated in Fig.\,\ref{fig:1D-sim}\,c), which shows the relation between quiver amplitude and decay length for $\delta \approx1$. Since the electrons start to experience the field decay already during their sub-cycle quiver motion, the ponderomotive potential transfer is increasingly non-adiabatic. As the effect is a consequence of the drift motion, it depends also on the laser pulse duration, which is most likely why it has not been recognized yet.\newline
Fig.\,\ref{fig:1D-sim}\,d) shows again the measured LEP values, now normalized by the ponderomotive potential ($\propto \mathcal{E}_0^2$ ) against the inverse quiver amplitude ($\propto \delta, \propto 1/\mathcal{E}_0 $) calculated from the incident intensity. Expressing the shift in terms of the natural length and energy scales allows better comparison with theory. Decreasing values of $1/l_{\mathrm{q,in}}$ first leads to an increase up to a maximum of around 15.5\,$\mathrm{U}_{\mathrm{p,in}}$ at $1/l_{\mathrm{q,in}}=4$, followed by a monotonic decrease.\newline
We now consider the effect of the finite laser pulse duration. Figure\,\ref{fig:1D-sim}\,c) shows the minimum electron energy for varying $\delta$ for emission during the central half-cycle of a Gaussian pulse with different pulse lengths $T_{\mathrm{FWHM}}$ and varying carrier-envelope phase (CEP $\in[-\pi,\pi]$)) as illustrated in the inset. For cw excitation ($T_{\mathrm{FWHM}}=\infty$, black line) the ponderomotive shift leading to the LEP decreases monotonically from $U_\mathrm{p}$ to 0 for decreasing $\delta$.\newline
For finite pulses and large values of $\delta$ the shift goes to zero. This can be understood by considering that the electrons do not fully leave the near-field before the field is switched off. The reduction of the ponderomotive potential difference is therefore more severe for shorter pulses. It is possible to estimate the reduction of the ponderomotive shift by solving an approximate equation for the drift motion for an electron without initial velocity in a time-dependent potential:
\begin{equation}
U_p(x,t)=U_{p0} \cdot \exp\Big(-\frac{2\,x}{l_f} - 4\,\mathrm{log}(2) \frac{t^2}{T_{\mathrm{FWHM}}^2}\Big),
\end{equation}
where $U_{p0}=e^2 E_{0}^2 / (4m \omega^2)$ is the amplitude of the ponderomotive potential at the surface. The gradient of the ponderomotive potential gives the force $F=-\nabla U_p(x,t)$, which leads to the equation of motion $\frac{\mathrm{d}v}{\mathrm{d}t}=-e\,F$. The evaluation of the expression is complicated by the $x(t)$-dependence. Assuming an electron initially at rest born at $t_0=0$ subject to a mean acceleration $\bar{a}$ (in the spirit of a Taylor-expansion of the equation of motion), we obtain for the position of the electron
$x(t)=0.5 \cdot \bar{a} \cdot(t-t_0)^2$. Substituting this expression into the equation of motion leads to the following result for the final velocity $\bar{v}_\mathrm{f}$:
\begin{equation}
\bar{v}_{ \mathrm{f}}=\frac{1}{2} \sqrt{\frac{\pi}{\bar{a}/l_f+4\,\mathrm{log}(2)/T_{\mathrm{FWHM}}^2}}.
\end{equation}
Now using $U_{p0}=l_q^2\cdot \frac{m \pi^2}{T_0^2}$, we can express the energy shift due to the time-dependent ponderomotive potential:
\begin{equation}
\label{Eq::shift_p}
\frac{\Delta E_\mathrm{p}}{U_p}=\frac{1}{\frac{2\bar{a}\cdot m \cdot l_f}{\pi \cdot U_{p0}}+\frac{8\,\mathrm{log}(2)}{\pi^3}\cdot\delta^{-2}\cdot(\frac{T_0}{T_{\mathrm{FWHM}}})^2}.
\end{equation}
In order to lead to the right behaviour for $T_{\mathrm{FWHM}}>>T_0$ and $\delta>>1$, the mean acceleration $\bar{a}$ has to be chosen such that the first term in the denominator equals unity. Similiar considerations have been done for photoemission in gases e.g. in Ref. \cite{Agostini87}. This function is shown as dashed line in Fig.\,\ref{fig:1D-sim}\,c). As can be seen, the expression describes quite well the decrease of the ponderomotive shift for increasing $\delta$.\newline
For smaller values of $\delta$ and finite pulses the LEP asymptotically reaches the values of the shift in a cw-field, which expresses the transition to the sub-cycle emission. Heuristically, this shift can be well approximated by the analytic function:
\begin{equation}
\label{Eq:shift_cw}
\frac{E_{\mathrm{shift,cw}}}{U_\mathrm{p}}=\frac{1}{1+a\,x^b}
\end{equation}
with the dimensionless variables a=3.8 and b=-1.3 obtained from a fit to the numerical result (black dotted line).\newline
The ponderomotive shift of the low-energy peak can be well approximated by multiplying Eq. \ref{Eq::shift_p} and Eq. \ref{Eq:shift_cw} such that $E_\mathrm{LEP}/U_\mathrm{p}=E_\mathrm{shift,p}/U_\mathrm{p}\cdot E_\mathrm{shift,cw}/U_\mathrm{p}$. This is shown as solid lines in Fig.\,\ref{fig:1D-sim}\,c), which yields excellent agreement with the values obtained from numerical simulations for different CEPs (shaded areas). This might be considered a consequence of the separability of sub-cycle motion and the drift motion, yet it is surprising that it works so well also in the intermediate regime. We have thus obtained an approximate analytic expression for the ponderomotive shift of the low-energy peak for arbitrary pulse lengths and adiabaticity parameters.\newline
Comparison of the experimental data in Fig.\,\ref{fig:1D-sim}\,d) with the analytical expression assuming a decay length of 4.5$\pm$0.5nm and a field enhancement factor of 5.5$\pm$0.5 (shaded area) yields satisfactory agreement. The seemingly linear dependence of the experimental LEP position on intensity in Fig.\,\ref{fig:exp_spectra}\,c) is due to the small changes of the normalized LEP shift close to the maximum of the curve in Fig.\,\ref{fig:1D-sim} d).The LEP shift could potentially be used to determine decay lengths and field enhancements directly from measurements. Since the Simple-Man's-Model and the strong-field approximation neglect effects such as image charge and Coulomb interaction of photoemitted electrons and as low-energy electrons are particularly sensitive to such effects, we do, however, not expect perfect agreement. In our experiments, the onset of the interaction of photoemitted electrons might be observable as a systematic reduction of the shift of the low-energy peak for increasing intensity (decreasing $1/l_\mathrm{q}$)\cite{Yanagisawa16}. We believe, that the non-adiabatic ponderomotive shift can provide a sensitive tool to measure such effects. \newline
For few-cycle pulses ($T_{\mathrm{FWHM}}/T_0\leqq 7.5$) the ponderomotive shift is only obtained for adiabaticity parameters below 50. For shorter pulses the maximum energy shift moves to lower $\delta$ and reaches smaller values. As can be seen from the shaded areas in Fig.\,\ref{fig:1D-sim}\,b), for shorter pulses also a significant CEP-dependence of the minimum energy is obtained. We point out that few-cycle pulses imply ponderomotive shifts can only be observed in the non-adiabatic regime. The exponential near-field decay has been chosen as we consider it to be the most generic form to describe a decaying near-field from arbitrary nanostructures and because it allows the analytical derivation of Eq. \ref{Eq::shift_p}. Nevertheless, we compared the resulting ponderomotive shift for continuous wave excitation to a field described by the component of a dipole parallel to the polarization direction, which is regularly used to model near-fields at the hemispherical apex of nanotips\cite{Novotny06, Ropers16}. Here the radiating dipole is assumed to be centered at a sphere of radius $R$, the electron propagation is started at the pole of the sphere and the field at the pole is normalized to $\mathcal{E}_0$. By identifying $R=2.5\cdot l_f$, which also yields good agreement of the two different field forms, the relative difference of the resulting low-energy peak position is less than 5$\%$ for adiabaticity parameters $>$1. Furthermore, using this relation we can estimate, from the above comparison of the analytic expression with the experimental data, the nanotip radius in the experiment to be around 10\,nm.
We point out that since the ponderomotive shift is fundamental in photoemission from nanostructures, it appeared, unrecognized, in previous theoretical\cite{Herink12,Ropers16,Lienau12,Lienau14,Dombi13} and experimental work (e.g.\cite{Ropers16}).\newline
\section{Numerical Simulation}
In the simple model, we only considered the LEP arising from emission by a single half-cycle and propagation in only one dimension. In order to investigate how the LEP persists for emission from a realistic nanotip and different half-cycles of a laser pulse we perform simulations similar to \cite{Lienau12,
Lienau14} with parameters fitting our experiment. The near-field around the nanotips is modelled using an analytic expression in the electrostatic approximation for a
hyperbolic tip shape \cite{Behr08} with a field decay length of roughly 4.5\,nm and a field enhancement factor of 6 consistent with numerical simulations of Maxwell's equations\cite{Thomas2015}. The electron emission probability is calculated using a simple Fowler-Nordheim equation\cite{Lienau12} assuming a work function of 5.5\,eV. After their emission electrons are propagated numerically in the near-fields using a fourth order Runge-Kutta approach. For simplicity a rescattering probability of unity for returning electrons is assumed. This will lead to an overestimation of rescattered electrons\cite{Krueger2012,Wachter2012}, however, since the low-energy region is dominated by direct electrons (see Fig.\,\ref{fig:1D-sim}\,a)), this will not affect the low-energy peak. Spatially, the emission sites considered in the simulation are restricted to regions on the tip apex where the obtained electron trajectories subtend angles approx. $\leq\pm 3.5^{\circ} $ (TOF acceptance angle) with respect to the tip axis.

\begin{figure}
	\centering
	\includegraphics[width=\linewidth]{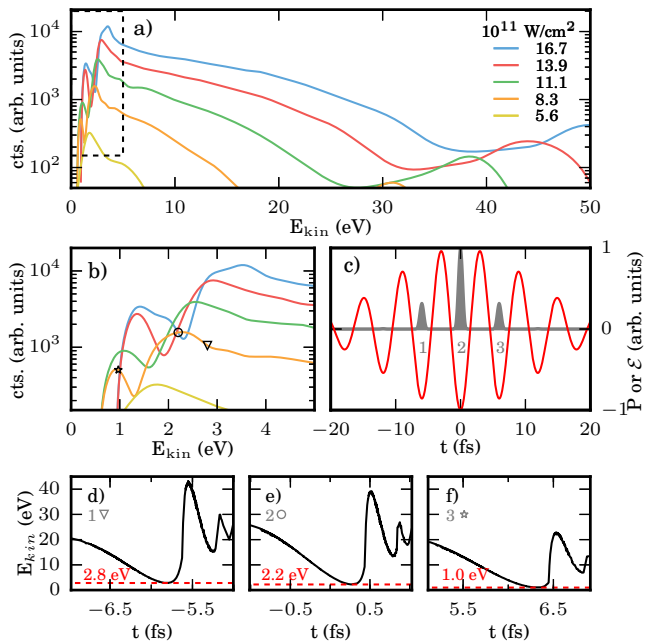}
	\caption{a) Simulated energy spectra of electron emission from a gold nanotip with an assumed near-field decay length of $\approx$4.5\,nm for indicated input
intensities. b) Extended view of the low-energy region, illustrating the shift of low-energy peaks. c) Electric field $\mathcal{E}$ of the laser pulse (solid line) used to simulate electron trajectories along with their half-cycle emission probability (shaded areas). d)-f) Final kinetic energies attained by each
electron emitted from the tip surface at different half-cycles of the laser pulse  (only shown for $ I=8.3\cdot10^{11}\,\mathrm{W/cm}^2 $). The turning points (marked above the dotted lines) contribute to the low-energy peaks in the spectra and are indicated in b).}
	\label{fig:simulated_spectra}
\end{figure}

The results of such simulations for different intensities are shown in Fig.\,\ref{fig:simulated_spectra}\,a). A prominent LEP is present and with increasing intensity a shoulder forms, similar to our experimental observations. An additional feature is given by an increase at higher energies after the drop of the shoulder, e.g. the increase beyond 32\,eV at the second highest intensity. This can be traced back to rescattered electrons, whose contribution is overestimated as mentioned above. The behavior of the LEP can be seen in Fig.\,\ref{fig:simulated_spectra}\,b). With increasing intensity, the peak height and position increases as expected from the simple 1D-model, indicating that even this model captured the main features in the experiments. Secondary peaks at lower energies appear as observed in the experiment. In order to resolve the origin of these secondary peaks, we compared the electron emission along the tip axis from different half-cycles of
the laser pulse as shown in Fig.\,\ref{fig:simulated_spectra}\,c), inspecting the relation between birth time and final energy shown in Fig.\,\ref{fig:simulated_spectra}\,d)-f) for $I$= $8.3\cdot10^{11}\,\mathrm{W/cm}^2$. Each half-cycle yields a specific LEP and the distinct peaks therefore correspond to different emission
times. The relative peak heights agree well with experimental observations and can be traced back to the different emission probabilities as illustrated in Fig.\,\ref{fig:simulated_spectra}\,c). For low intensities, the contribution of the first cycle is hidden in the main peak as illustrated in Fig.\,\ref{fig:simulated_spectra}\,b). The LEP shift compared to the cutoff is significantly higher and the change of the height of the main low-energy peak with increasing intensity is smaller than observed experimentally, and we could not find suitable parameters where they agreed. We attribute this mainly to shortcomings of our model which neglects effects such as image charge interaction as well as details of the emission process. Overcoming this limitations would require approaches on the many-body quantum level such as time-dependent density-functional theory (TDDFT)\cite{Wachter2012}, which is to our knowledge out of reach for 3d-nanogeometries. The overall good agreement between the simulations and experiments, nevertheless, permits to clearly identify the non-adiabatic ponderomotive shift as origin for the LEP shifts with intensity.

\section{Conclusion and Outlook}
In conclusion, we have shown an intensity-dependent study revealing the up-shift of a low-energy peak in the photoemission from a metallic nanotip. We identified the adiabaticity parameter and the pulse length as the relevant parameters and derived a simple analytic expression for the ponderomotive shift. The results show that small near-field decay lengths inhibit the separation of sub-cycle quiver motion and cycle-integrated drift motion, and the ponderomotive shift becomes non-adiabatic. Importantly for few-cycle pulses, the ponderomotive shift is only observable for small adiabaticity parameters and therefore generally non-adiabatic. Simulations taking into account a realistic emission geometry could assign additional small peaks in the spectra to photoemission from different half-cycles of the laser pulse. The findings here about the non-adiabatic ponderomotive shift are an important, previously not studied effect, completing the understanding of the role of the inhomogeneous field in strong-field photoemission from nanostructures. Since the LEP shift relies on low-energy electrons, it could possibly be used as a sensitive tool to measure effects that go beyond the Strong-Field-Approximation such as image charge interaction and Coulomb interaction. The effect might also be used to control the energy of the most dominant contribution in nanotip photoemission, which could for instance be useful in generating energy-tuneable electrons for inducing ultrafast molecular reactions.

We are grateful to F. Krausz for the use of specialized equipment, and we acknowledge fruitful discussions with B. Ahn, D. Kim, T. Fennel, T. Zimmermann, and A. Landsman. We are grateful for support by the Max Planck Society and the DFG through LMUexcellent, SPP1840 and the Cluster of Excellence: Munich Centre for Advanced Photonics
(MAP). We acknowledge support from the European Union (EU) via the European Research Council (ERC) grants ATTOCO and NearFieldAtto. M.F.C. was supported by the project ELI-Extreme Light Infrastructure-phase2 (Grant No. CZ.02.1.01/0.0/0.0/15\_008/0000162) from the European Regional Development Fund.

\bibliographystyle{apsrev4-1}
\bibliography{Literature}

\end{document}